\documentclass[lettersize,journal]{IEEEtran}
\usepackage{amsmath,amsfonts}
\usepackage{algorithmic}
\usepackage{algorithm}
\usepackage{array}
\usepackage[caption=false,font=normalsize,labelfont=sf,textfont=sf]{subfig}
\usepackage{textcomp}
\usepackage{stfloats}
\usepackage{url}
\usepackage{verbatim}
\usepackage{graphicx}
\usepackage{cite}
\usepackage{multirow}
\begin{document}

\title{XEmoGPT: An Explainable Multimodal Emotion Recognition Framework with Cue-Level Perception and Reasoning}

\author{
Hanwen Zhang,
Yao Liu,
Peiyuan Jiang,
Lang Junjie,
Xie Jun,
Yihui He,
Yajiao Deng,
Siyu Du,
and Qiao Liu
\thanks{Hanwen Zhang, Yao Liu, Peiyuan Jiang, Yihui He, Yajiao Deng, Siyu Du and Qiao Liu are with the School of Computer Science and Engineering, University of Electronic Science and Technology of China, Chengdu, Sichuan, China.}
\thanks{Lang Junjie and Xie Jun are with 54th Research Institute, China Electronics Technology Group Corporation, Shijiazhuang, Hebei, China.}
\thanks{This work was supported by the National Natural Science Foundation of China U22B2061 and the Natural Science Foundation of Sichuan, China 2024NSFSC0496.}
\thanks{Corresponding authors: Yao Liu and Qiao Liu.}
}

\markboth{Journal of \LaTeX\ Class Files,~Vol.~14, No.~8, August~2021}%
{Shell \MakeLowercase{\textit{et al.}}: A Sample Article Using IEEEtran.cls for IEEE Journals}


\maketitle

\begin{abstract}
Explainable Multimodal Emotion Recognition plays a crucial role in applications such as human-computer interaction and social media analytics. However, current approaches struggle with cue-level perception and reasoning due to two main challenges: 1) general-purpose modality encoders are pretrained to capture global structures and general semantics rather than fine-grained emotional cues, resulting in limited sensitivity to emotional signals; and 2) available datasets usually involve a trade-off between annotation quality and scale, which leads to insufficient supervision for emotional cues and ultimately limits cue-level reasoning. Moreover, existing evaluation metrics are inadequate for assessing cue-level reasoning performance. To address these challenges, we propose e\textbf{X}plainable \textbf{Emo}tion \textbf{GPT} (XEmoGPT), a novel EMER framework capable of both perceiving and reasoning over emotional cues. It incorporates two specialized modules: the Video Emotional Cue Bridge (VECB) and the Audio Emotional Cue Bridge (AECB), which enhance the video and audio encoders through carefully designed tasks for fine-grained emotional cue perception. To further support cue-level reasoning, we construct a large-scale dataset, EmoCue, designed to teach XEmoGPT how to reason over multimodal emotional cues. In addition, we introduce EmoCue-360, an automated metric that extracts and matches emotional cues using semantic similarity, and release EmoCue-Eval, a benchmark of 400 expert-annotated samples covering diverse emotional scenarios. Experimental results show that XEmoGPT achieves strong performance in both emotional cue perception and reasoning.
\end{abstract}

\begin{IEEEkeywords}
Multimodal Large Language Models, Explainable Multimodal Emotion Recognition, Model Explainability.
\end{IEEEkeywords}

\section{Introduction}

\begin{figure}[t]
\centering
\includegraphics[width=\linewidth]{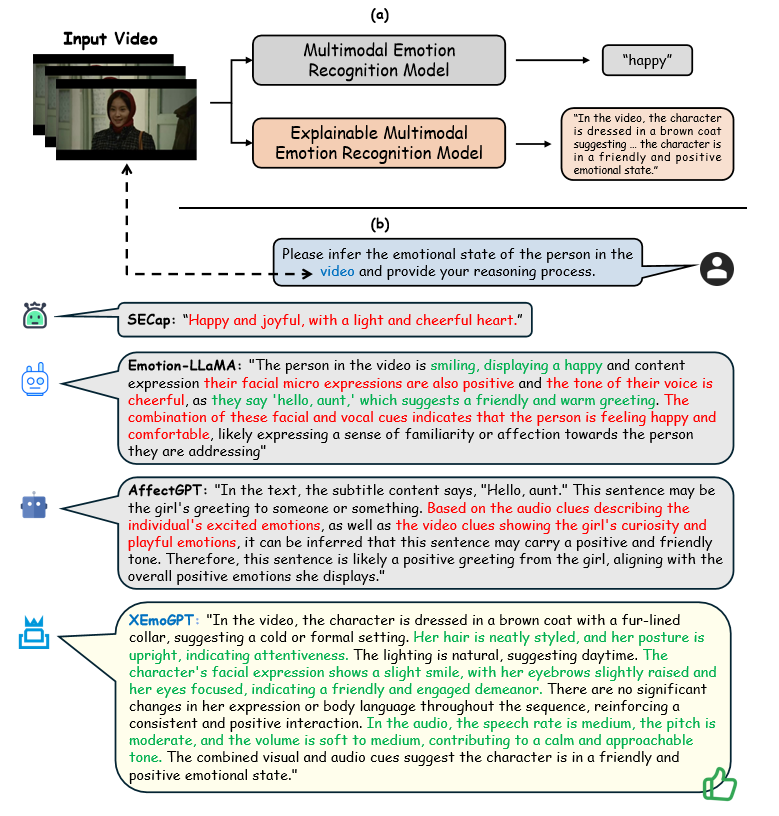}
\caption{(a) Comparison between Multimodal Emotion Recognition models and Explainable Multimodal Emotion Recognition models. (b) Comparison between XEmoGPT and other Emotional MLLMs: Green/red text indicates emotional predictions with/without explicit cue-level explanations.}
\label{fig:1}
\end{figure}


\IEEEPARstart{E}{motion-aware} technologies play a pivotal role in a wide range of real-world applications, including human–computer interaction~\cite{preece1994human}, social media analysis~\cite{batrinca2015social}, and social robotics~\cite{breazeal2016social}. As these systems increasingly participate in decision-making processes that directly affect human users, merely predicting emotional labels is no longer sufficient. Instead, there is a growing demand for models that can explicitly justify their predictions with transparent and explainable evidence, enabling users to understand why a certain emotional state is inferred. This requirement has driven the paradigm shift from traditional Multimodal Emotion Recognition (MER)~\cite{ahmed2023systematic}, which focuses on outcome-level emotion classification, toward Explainable Multimodal Emotion Recognition (EMER)~\cite{lian2023explainable}, which emphasizes evidence-grounded emotional reasoning. As shown in Figure~\ref{fig:1}(a), EMER goes beyond merely predicting emotional labels and aims to uncover the underlying perceptual and reasoning processes that support emotional inference.

With the rapid development and widespread adoption of Multimodal Large Language Models (MLLMs), recent EMER approaches have leveraged their generative capacity to produce emotional descriptions~\cite{wu2023multimodal, xie2024emovit, liang-etal-2024-aligncap, lian2025affectgptr1leveragingreinforcementlearning, zhang2025rethinkingfacialexpressionrecognition}, such as models like SECap~\cite{xu2024secap}, Emotion-LLaMA~\cite{cheng2024emotion}, and AffectGPT~\cite{lian2025affectgpt}. Built upon instruction-tuned LLaMA-2~\cite{touvron2023llama}, Emotion-LLaMA first encodes audio, facial and contextual visual cues with modality-specific experts, projects them into a shared token space, and then autoregressively generates an emotion label together with a free-form textual explanation; AffectGPT extends this pipeline with a lightweight pre-fusion module, such as a Q-Former or attention-based pooling, placed between the unimodal encoders and the Qwen-2.5 LLM~\cite{qwen2025qwen25technicalreport}, so that cross-modal interactions are effectively distilled prior to generating detailed emotional descriptions. However, despite their promising performance, existing methods largely produce result-oriented descriptions that resemble emotion paraphrasing rather than genuine explanations. As shown in Figure~\ref{fig:1}(b), such descriptions usually state the predicted emotion and, at most, refer to vague “cues” without specifying the concrete multimodal evidence supporting the inference. Consequently, the explanations lack transparency and are difficult to verify at a fine-grained level.


To address this issue, we introduce the concept of \textit{emotional cues} as explicit, modality-specific evidential units that bridge low-level perception and high-level emotional reasoning. An emotional cue corresponds to an observable fact grounded in one or more modalities—such as facial expressions, body posture, vocal tone, or contextual objects—from which an emotional state can be logically inferred. For instance, raised eyebrows and a gentle smile constitute visual cues that suggest happiness, while a lowered head posture combined with a tense facial expression may indicate sadness. By decomposing emotional reasoning into a structured hierarchy consisting of perceptual observation, cue identification, reasoning, and emotional inference, emotional cues emerge as explainable intermediate variables that explicitly connect low-level sensory signals with high-level emotional conclusions.

Despite their success in general multimodal understanding, existing MLLM-based EMER approaches face two fundamental challenges that hinder cue-level reasoning: (1) The pretraining objectives of general-purpose modality encoders are primarily designed to capture global structures and general semantic representations rather than the fine-grained emotional cues, which consequently leads to limited sensitivity to emotional signals, as shown in Table~\ref{tab:exp2}. (2) As detailed in Section~\ref{sec:training_dataset}, existing datasets often present a trade-off between annotation quality and scale, where large datasets lack precise emotional supervision and smaller high-quality ones are limited in data scale, leading to insufficient cue-level guidance for emotional reasoning.

To enhance emotional cue perception, we propose XEmoGPT, which integrates two modules: the Video Emotional Cue Bridge (VECB) and the Audio Emotional Cue Bridge (AECB). VECB enhances the visual encoder's ability to perceive temporal visual emotional cues via positional encoding and self-attention, and is further trained on auxiliary tasks including Contrastive Video Emotional Cue Alignment, Frame Temporal Discrimination, and Masked Frame Modeling. AECB enhances the audio encoder’s sensitivity to auditory emotional cues through training on the task of Contrastive Audio Emotional Cue Alignment.

To enable fine-grained emotional reasoning, we introduce the EmoCue dataset, which provides cue-level supervision for emotional reasoning. It consists of: (1) EmoCue-Instruct, which refines descriptions in MER-Caption+ via a \textit{model-led}, \textit{human-assisted} strategy~\cite{lian2025affectgpt} to inject explicit modality-specific cues; and (2) EmoCue-ShortCaption, which leverages MLLMs to generate concise emotional cue annotations.

Moreover, current evaluation metrics for EMER largely rely on prompt-based scoring~\cite{lian2023explainable}, which fails to accurately measure the overlap between model-generated and human-annotated texts at the level of emotional cues. To address this limitation, we introduce EmoCue-360, a novel automatic evaluation metric that extracts visual, auditory, and global emotional cues from generated descriptions and computes cue-level F1 scores based on semantic similarity. To support systematic evaluation, we further release EmoCue-Eval, a human-annotated benchmark containing 400 high-quality samples spanning diverse scenarios and emotional states. To the best of our knowledge, it is the largest benchmark dedicated to EMER. Our key contributions are summarized as follows:\begin{itemize}
    \item We propose XEmoGPT with VECB and AECB modules to enhance emotional cue perception via modality-specific cue alignment. Extensive experiments show that XEmoGPT achieves state-of-the-art performance on both conventional generation metrics and EmoCue-360, consistently demonstrating its robust and comprehensive capabilities in both emotional cue perception and reasoning.
    \item We construct the EmoCue dataset, which comprises two parts: EmoCue-Instruct, containing 30k+ samples with explicit annotations of emotional states and modality-specific cues (visual, audio, and global); and EmoCue-ShortCaption, containing 40k+ model-generated short emotional cues. 
    \item We design EmoCue-360, an automated evaluation metric that calculates F1 scores at the cue level. We also release EmoCue-Eval, the largest human-annotated EMER benchmark with 400 samples for systematic evaluation.
\end{itemize}

\section{Related Work}
\label{sec:related_work}

\subsection{Multimodal Large Language Models}

Multimodal Large Language Models (MLLMs) have gained strong traction in tasks such as visual question answering~\cite{antol2015vqa}, image-to-text generation~\cite{li2023evaluating}, and cross-modal reasoning~\cite{liu2024mmbench}. A typical MLLMs architecture is composed of three core modules: a pre-trained modality encoder, a pre-trained large language model (LLM), and a modality interface that bridges them~\cite{wu2023multimodal}. Drawing an analogy to human cognition, the modality encoder, such as CLIP-ViT~\cite{radford2021learning} or EVA-CLIP~\cite{sun2023eva} for images and CLAP~\cite{wu2023large} for audio, acts as the sensory system that perceives and preprocesses raw signals into compact representations. The LLM, often instantiated with models like LLaMA~\cite{touvron2023llama} or Vicuna~\cite{zheng2023judging}, serves as the central brain responsible for reasoning and response generation. The modality interface aligns heterogeneous modalities into a unified representation space understandable by the LLM. It can be implemented via learnable connectors, including the Q-Former used in BLIP-2\cite{li2023blip} and InstructBLIP~\cite{dai2023instructblip}, or MLP projectors adopted in LLaVA~\cite{li2024llava}, as well as through expert models that convert non-text inputs into language. However, existing MLLMs exhibit limited capabilities in affective scenarios~\cite{hu2025emobench, lian2024merbench}. Even advanced models like GPT-4V~\cite{LIAN2024102367} struggle to recognize audio emotional cues or subtle facial micro-expressions, mainly due to the gap between general pretraining objectives and EMER task requirements.

\subsection{Explainable Multimodal Emotion Recognition}


Explainable Multimodal Emotion Recognition (EMER) task aims to identify emotional states from video, audio, and text modalities, while simultaneously providing explicit and explainable justifications for the predicted emotions~\cite{lian2023explainable}. Recent advances in affective computing have explored emotion-specific fine-tuning of MLLMs~\cite{xie2024emovit, liang-etal-2024-aligncap}, including models such as SECap~\cite{xu2024secap}, Emotion-LLaMA~\cite{cheng2024emotion}, and AffectGPT~\cite{lian2025affectgpt}, thereby offering a promising research direction for EMER. However, these methods often generate emotion-centric descriptions that emphasize the inferred emotional states from visual or auditory cues, without explicitly specifying the concrete multimodal evidence underlying the inferences.

Moreover, existing automated evaluation metrics for EMER remain limited in both reliability and interpretability. The prevailing approach relies on "Prompt-based Scoring"~\cite{lian2023explainable}: feeding both model-generated text and human-annotated text into a LLM along with carefully designed prompts to generate scores. Despite its simplicity, this metric has certain limitations: its scoring process can be influenced by prompt design and the inherent subjectivity of LLM-generated outputs, which may affect the objectivity and consistency of evaluating alignment and information overlap at the emotional cue level. These factors can, to some extent, challenge the reproducibility and reliability of the evaluation results.

\begin{figure*}[t]
\centering
\includegraphics[width=\textwidth]{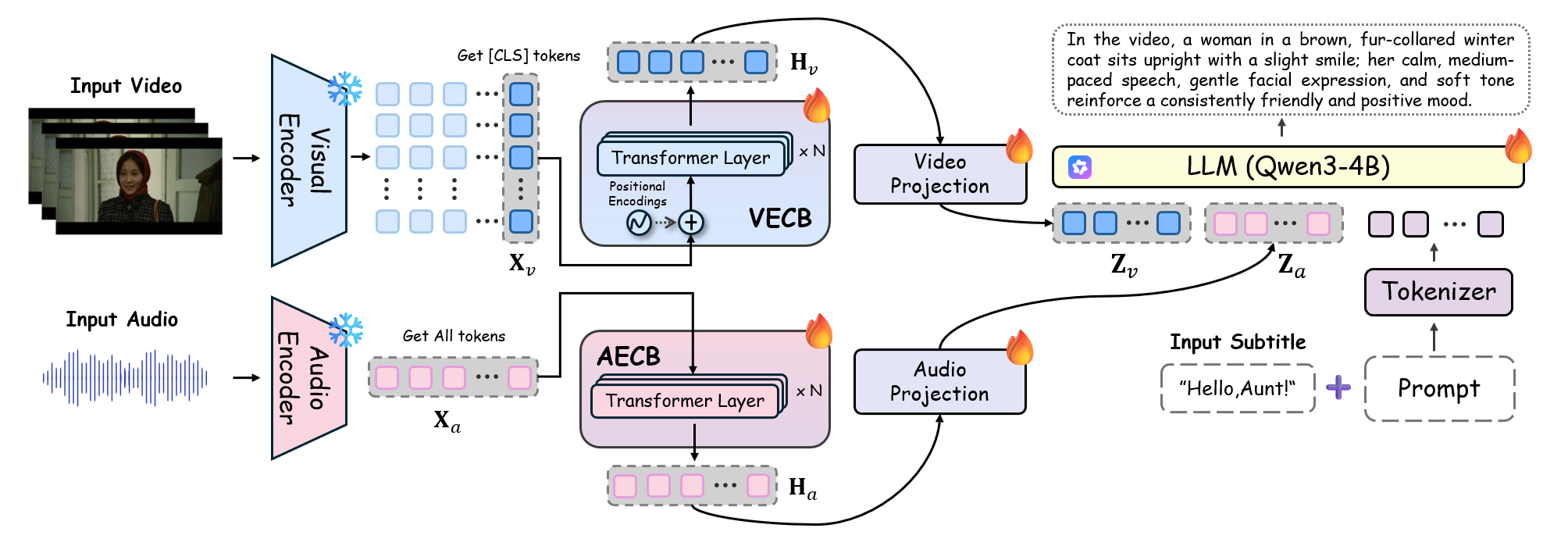}
\caption{Architecture of XEmoGPT: It integrates visual, auditory, and textual information to generate a description containing both visual and auditory emotional cues. The VECB and AECB modules primarily serve to enhance the emotional cue perception capabilities of the modality encoders.}
\label{fig:2}
\end{figure*}

\section{Methodology}

This section presents the architecture of XEmoGPT, followed by design details of the Video Emotional Cue Bridge and the Audio Emotional Cue Bridge, and concludes with the training process.

\subsection{Model Architecture}

XEmoGPT consists of four components that are consistent with the conventional MLLM architecture, namely the visual encoder, audio encoder, projection layer, and the LLM, as shown in Figure~\ref{fig:2}. To enhance emotional cue perception capabilities, two novel modules have been introduced: the Video Emotional Cue Bridge (VECB) and the Audio Emotional Cue Bridge (AECB). 

First, we use the frozen CLIP-ViT~\cite{radford2021learning} as the visual encoder and the frozen HuBERT~\cite{hsu2021hubert} as the audio encoder to convert video frames and audio into a set of visual embeddings \(\mathbf{X}_v \in \mathbb{R}^{l_v \times d_v}\) and audio embeddings \(\mathbf{X}_a \in \mathbb{R}^{l_a \times d_a}\), where \(l_v\) is the length of video frames, \(l_a\) is the length of the audio embeddings, and \(d_v\) and \(d_a\) are the dimensions of the visual and audio embeddings, respectively. Next, we utilize the VECB to transform the visual embeddings \(\mathbf{X}_v\) into visual emotion embeddings \(\mathbf{H}_v \in \mathbb{R}^{l_v \times d_v}\), and the AECB to transform the audio embeddings \(\mathbf{X}_a\) into audio emotion embeddings \(\mathbf{H}_a \in \mathbb{R}^{l_a \times d_a}\). Further, we use the visual projection layer and the audio projection layer to convert \(\mathbf{H}_v\) and \(\mathbf{H}_a\) into visual emotion tokens \(\mathbf{Z}_v \in \mathbb{R}^{l_v \times d}\) and audio emotion tokens \(\mathbf{Z}_a \in \mathbb{R}^{l_a \times d}\), respectively. Finally, the visual emotion tokens \(\mathbf{Z}_v\), audio emotion tokens \(\mathbf{Z}_a\), the video subtitle \(\mathbf{T}\), and the prompt \(\mathbf{P}\) are fed into the LLM to generate the final response \(\hat{\mathbf{R}} = \text{LLM}(\mathbf{Z}_v, \mathbf{Z}_a, \mathbf{T}, \mathbf{P})\).

\subsubsection{Video Emotional Cue Bridge}

To enable distinguishing the order of frames at different time steps, the VECB adds a set of learnable positional encodings to the visual embeddings. Then, to cross-fusion the visual embeddings \(\mathbf{X}_v\) in the temporal dimension, we introduce a Transformer within the VECB module, ultimately obtaining the visual emotion embeddings \(\mathbf{H}_v\). Formally, this process can be represented as
\begin{equation}
    \mathbf{H}_v = \text{Transformer}(\mathbf{X}_v + \text{PE}(\mathbf{X}_v))
\end{equation}
where $\text{PE}(\cdot)$ represents the positional encoding.

\subsubsection{Audio Emotional Cue Bridge}

Since HuBERT can output fine-grained, time-step aligned audio representations and has already introduced positional encodings internally to model the temporal information of audio, the AECB module does not need to add additional positional encodings. Instead, it directly uses a Transformer to perform cross-fusion of the audio embeddings \(\mathbf{X}_a\) in the temporal dimension. This process can be represented as
\begin{equation}
    \mathbf{H}_a = \text{Transformer}(\mathbf{X}_a)
\end{equation}

\begin{figure*}[t]
\centering
\includegraphics[width=\textwidth]{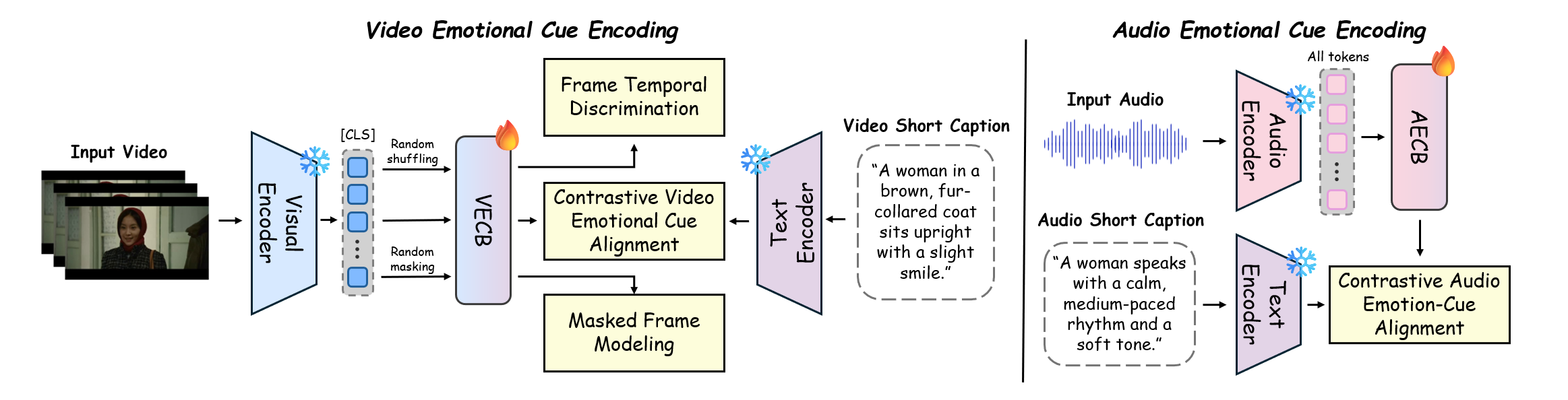}
\caption{Training process of VECB and AECB: The VECB module is trained with three auxiliary tasks: Contrastive Video Emotional Cue Alignment, Frame Temporal Discrimination, and Masked Frame Modeling. The AECB module is trained with the Contrastive Audio Emotional Cue Alignment task.}
\label{fig:3}
\end{figure*}

\subsection{Video Emotional Cue Encoding}

To enhance cue-level perception of the pre-trained visual encoder and capture temporal information in the video frame sequence, we jointly train the VECB module with multiple auxiliary tasks, including the Contrastive Video Emotional Cue Alignment task, the Frame Temporal Discrimination task, and the Masked Frame Modeling task, as shown in Figure~\ref{fig:3}.

\subsubsection{Contrastive Video Emotional Cue Alignment}

To bridge the gap between the visual embeddings and emotional cues, we introduce a contrastive learning objective between the output of VECB and the emotional cue representations. Specifically, we first pooling the VECB output $\mathbf{H}_v$ across the temporal dimension to obtain a global video representation $\mathbf{h}_v$. Meanwhile, the emotional cue is encoded using a frozen CLIP text encoder to obtain $\mathbf{h}_t$. To align the video and emotional cue representations, we adopt the InfoNCE loss~\cite{oord2018representation}, a widely used contrastive learning objective. It encourages paired video and emotional cue embeddings to be close in the representation space while pushing apart unpaired samples:
\begin{equation}
\label{eq3}
\mathcal{L}_{\text{cl}}^{\text{vt}} = - \log \frac{\exp(\text{sim}(\mathbf{h}_v, \mathbf{h}_t)/\tau)}{\sum_{j=1}^{N} \exp(\text{sim}(\mathbf{h}_v, \mathbf{h}_t^{(j)})/\tau)}
\end{equation}
where $\text{sim}(\cdot, \cdot)$ denotes cosine similarity, $\tau$ is a learnable temperature parameter, and $N$ is the batch size.

\subsubsection{Frame Temporal Discrimination}

To enable VECB to understand temporal dynamics and the order of frames, we propose a frame temporal discrimination task. Given a pair of video clips, we sample frame sequences and apply temporal shuffling. The model is then asked to predict whether the sequence is in correct temporal order. A simple binary classifier is applied on top of the global video representation $\mathbf{h}_v$ obtained via pooling over the VECB output:
\begin{equation}
\hat{y} = \text{MLP}(\mathbf{h}_v), \quad \mathcal{L}_{\text{temp}} = \text{BCE}(\hat{y}, y)
\end{equation}
where \( y \in \{0,1\} \) indicates if the frame sequence is ordered.

\subsubsection{Masked Frame Modeling}

To enhance the contextual understanding of frames, we adopt a masked frame modeling strategy, similar to masked language modeling in BERT~\cite{devlin2019bert}. A subset of visual embeddings in $\mathbf{X}_v$ is randomly masked and replaced with a learnable token. The VECB is then trained to reconstruct the original frame embeddings at the masked positions using a frame reconstruction loss:
\begin{equation}
\mathcal{L}_{\text{mfm}} = \sum_{i \in \mathcal{M}} \left\| \mathbf{\hat{x}}_v^{(i)} - \mathbf{x}_v^{(i)} \right\|_2^2
\end{equation}
where $\mathcal{M}$ denotes the set of masked positions, and $\mathbf{\hat{x}}_v^{(i)}$ is the predicted embedding at position $i$.

\subsection{Audio Emotional Cue Encoding}

Since HuBERT is pre-trained on large-scale unsupervised audio data, its representations possess strong contextual modeling capabilities within the audio space. However, due to the absence of large-scale audio-text paired training, there exists a significant semantic gap between its feature space and the semantic space of emotional cues. Therefore, the training objective of AECB is to enhance the emotional cue perception capability of the audio encoder by aligning its output features with the semantic space of emotional cues, as shown in Figure~\ref{fig:3}.

\subsubsection{Contrastive Audio Emotion-Cue Alignment}

Firstly, we pool the audio emotion embeddings \(\mathbf{H}_a\) temporally to obtain a global audio representation \(\mathbf{h}_a\). The emotion caption is encoded by a frozen LaBSE~\cite{feng2020language} to get \(\mathbf{h}_t\). LaBSE excels in multilingual, multi-domain sentence similarity, making it suitable for aligning sentence-level emotional cues. We still adopt the InfoNCE loss to align audio and emotional cues:
\begin{equation}
\mathcal{L}_{\text{cl}}^{\text{at}} = -\log \frac{\exp(\text{sim}(\mathbf{h}_a, \mathbf{h}_t)/\tau)}{\sum_{j=1}^{N} \exp(\text{sim}(\mathbf{h}_a, \mathbf{h}_t^{(j)})/\tau)}
\end{equation}
where symbols have the same meanings as in Equation~\ref{eq3}.

\subsection{Training Process}

The training process consists of three stages. The first stage focuses on training the VECB module to obtain video emotional cue representations. The second stage trains the AECB module to obtain audio emotional cue representations. The final stage aligns these multimodal representations with the text representation space of the LLM. The datasets used during training are listed in Table~\ref{tab:met1}.

\subsubsection{Stage1: Training VECB}

In this stage, both the visual encoder and text encoder are frozen, while the VECB module is randomly initialized. The training objective is defined as:
\begin{equation}
    \mathcal{L}_{\text{VECB}} = \mathcal{L}_{\text{cl}}^{\text{vt}} + w_1 \cdot \mathcal{L}_{\text{temp}} + w_2 \cdot \mathcal{L}_{\text{mfm}}
\end{equation}
where \(w_1\) and \(w_2\) are the weighting coefficients for the Frame Temporal Discrimination loss and the Masked Frame Modeling loss, respectively.

\subsubsection{Stage2: Training AECB}

In this stage, both the audio encoder and text encoder are frozen, and the AECB module is randomly initialized. The training objective is defined as:
\begin{equation}
    \mathcal{L}_{\text{AECB}} = \mathcal{L}_{\text{cl}}^{\text{at}}
\end{equation}

\subsubsection{Stage3: Emotional Instruction Fine-tuning}

During this stage, the parameters of the CLIP and HuBERT encoders remain frozen. For the LLM, we apply LoRA~\cite{hu2022lora} to enable efficient fine-tuning, allowing it to adapt to the EMER task with minimal computational overhead. In contrast, the VECB and AECB modules, along with the audio and visual projection layers, are fully trainable.

To enhance the model’s ability to reason over modality-specific cues, we design three types of instruction–response pairs during training: (1) visual-only, (2) audio-only, and (3) multimodal (both visual and audio). During optimization, we adopt a targeted gradient control strategy: for visual-only samples, gradients in the audio branch (including AECB and its projection layer) are blocked; conversely, for audio-only samples, gradients in the visual branch are blocked. This selective gradient flow enforces modality-specific inductive bias, enabling the model to focus on the relevant modality during inference based on user instructions, while avoiding cross-modal confusion.

\section{Training Dataset}
\label{sec:training_dataset}

\begin{table}
  \centering
  \renewcommand{\arraystretch}{1.5}
  \caption{Training datasets used in the three training stages of XEmoGPT. Stage 1 and Stage 2 use video/audio-text datasets; and Stage 3 uses multimodal instruction-tuning datasets for aligning emotional reasoning.}
  \begin{tabular}{c|>{\centering\arraybackslash}m{17em}|c}
    \hline
    Stage & Datasets & Size \\
    \hline
    1 & MAFW~\cite{liu2022mafw}, DFEW~\cite{jiang2020dfew}, MER2025 Track2~\cite{lian2025mer}, EmoCue-ShortCaption & 0.9M \\
    \hline
    2 & SpeechCraft-Zhvoice~\cite{jin2024speechcraft} & 10.2M \\
    \hline
    3 & MAFW~\cite{liu2022mafw}, EmoCue, MERCaption+~\cite{lian2025affectgpt}, SpeechCraft-50k~\cite{jin2024speechcraft} & 1.6M \\
    \hline
  \end{tabular}
  \label{tab:met1}
\end{table}

Currently, datasets for the EMER task face a trade-off between sample size and annotation quality. Existing resources exhibit complementary strengths and limitations. MERR-Coarse-Grained~\cite{cheng2024emotion} provides a large-scale dataset (number of samples N = 28,618) with coarse-grained annotations, while MERR-Fine-Grained~\cite{cheng2024emotion} offers more detailed annotations but at a smaller scale (N = 4,487). MER-Caption+~\cite{lian2025affectgpt} achieves a more balanced compromise between sample size (N = 31,327) and annotation quality, though it primarily focuses on textual emotion descriptions without explicitly labeling visual or auditory cues.

To overcome these limitations, We propose EmoCue, a new dataset composed of two parts: EmoCue-Instruct (N = 31,327) and EmoCue-ShortCaption (N = 40,683). 

EmoCue-Instruct builds upon MER-Caption+ by rewriting its annotations through a \textit{model-led}, \textit{human-assisted} strategy: (1) MLLMs are prompted with MER2025~\cite{lian2025mer} emotion labels to infer modality-specific emotional cues; (2) \texttt{deepseek-chat}~\cite{deepseekai2025deepseekr1incentivizingreasoningcapability, deepseekai2025deepseekv32pushingfrontieropen} consolidates these cues, ensuring consistency and resolving conflicts; (3) Experts randomly selected 300 samples and manually scored them to assess the quality of the dataset, as shown in Section~\ref{sec:eval_emocue_inst}. 

EmoCue-ShortCaption, built on DFEW~\cite{jiang2020dfew} and MER-Caption+, leverages MLLMs to generate concise emotional cues without human involvement.

\begin{figure*}[t]
\centering
\includegraphics[width=\textwidth]{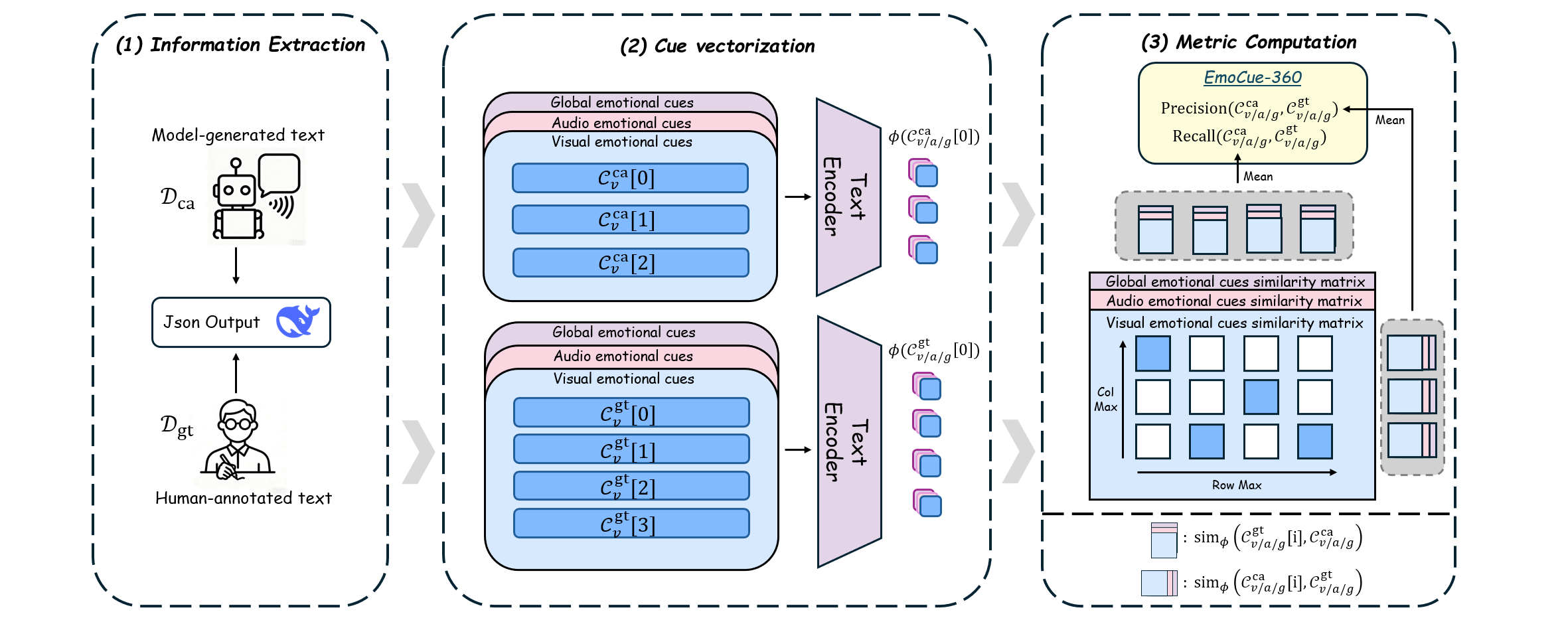}
\caption{Computation pipeline of the EmoCue-360 metric: Information Extraction, Cue Vectorization, and Metric Computation. The mathematical formulas used correspond to those in Section~\ref{sec:eval}.}
\label{fig:4}
\end{figure*}

\section{Evaluation Metric and Dataset}
\label{sec:eval}

As discussed in Section~\ref{sec:related_work}, current evaluation methods for the EMER task remain limited. To address these limitations, inspired by~\cite{Ye_2025_CVPR}, we propose a novel automated evaluation metric, EmoCue-360, which precisely quantifies the overlap between model-generated and human-annotated texts at the level of emotional cues.

\subsection{Task Formulation}

We first provide a more detailed definition of the EMER task: given a video segment, the model is required to generate an explainable emotional description \( \mathcal{D} \), which should consist of three parts of information at the semantic level:
\begin{equation}
    \mathcal{D} = \{ \mathcal{C}_v, \mathcal{C}_a, \mathcal{C}_s \}
\end{equation}
where \( \mathcal{C}_v \), \( \mathcal{C}_a \), and \( \mathcal{C}_s \) represent the sets of visual, audio, and global emotional cues, respectively. Each cue set is composed of several atomic emotional cues \( c \). These atomic cues can be regarded as the smallest discernible, semantically independent units in the text. In theory, any natural language text can be decomposed into such atomic cues.~\cite{brachman2004knowledge}

Given a pair of model-generated text \( \mathcal{D}_{\text{ca}} \) and human-annotated text \( \mathcal{D}_{\text{gt}} \), each dimension of a high-quality \( \mathcal{D}_{\text{ca}} \) should satisfy two conditions:
\begin{itemize}
    \item \textbf{Completeness}: any atomic cue that appears in a dimension of \( \mathcal{D}_{gt} \) should have a semantically identical atomic cue appearing in the corresponding dimension of \( \mathcal{D}_{ca} \).
    \item \textbf{Correctness}: any atomic cue that appears in a dimension of \( \mathcal{D}_{ca} \) should have a semantically identical atomic cue appearing in the corresponding dimension of \( \mathcal{D}_{gt} \).
\end{itemize}

\subsection{Metric Computation}

The computation process of the EmoCue-360 metric is illustrated in Figure~\ref{fig:4}. Given a pair of model-generated text \( \mathcal{D}_{\text{ca}} \) and human-annotated text \( \mathcal{D}_{\text{gt}} \), we first need to extract the sets of emotional cues from the texts in the three dimensions mentioned above. To address the open-ended text cue extraction problem, we can leverage the strong capability of LLM in information extraction tasks~\cite{Xu2023LargeLM} to automatically decompose open-ended text into refined cue lists.

Next, for each dimension of \( \mathcal{D}_{\text{ca}} \) and \( \mathcal{D}_{\text{gt}} \), we can calculate the precision and recall. Taking the visual emotional cues as an example, the set of visual emotional cues in the model-generated text is \( \mathcal{C}_v^{\text{ca}} \), and the set of visual emotional cues in the human-annotated text is \( \mathcal{C}_v^{\text{gt}} \). Thus, the precision and recall are defined as:
\begin{equation}
    \text{Precision}(\mathcal{C}_v^{\text{ca}}, \mathcal{C}_v^{\text{gt}}) = \frac{1}{|\mathcal{C}_v^{\text{ca}}|} \sum_{i=1}^{|\mathcal{C}_v^{\text{ca}}|} \text{sim}_\phi(\mathcal{C}_v^{\text{ca}}[i], \mathcal{C}_v^{\text{gt}})
\end{equation}
\begin{equation}
    \text{Recall}(\mathcal{C}_v^{\text{ca}}, \mathcal{C}_v^{\text{gt}}) = \frac{1}{|\mathcal{C}_v^{\text{gt}}|} \sum_{i=1}^{|\mathcal{C}_v^{\text{gt}}|} \text{sim}_\phi(\mathcal{C}_v^{\text{gt}}[i], \mathcal{C}_v^{\text{ca}})
\end{equation}
where \( \text{sim}_{\phi}(\mathcal{C}[i], \mathcal{C}') = \max_{\mathcal{C}'[j] \in \mathcal{C}'} \cos\left( \phi(\mathcal{C}[i]), \phi(\mathcal{C}'[j]) \right) \) denotes the similarity between the cue \( \mathcal{C}[i] \) and the cue set \( \mathcal{C}' \), \( \phi(\cdot) \) is a sentence-level text encoder that maps textual cues into vector representations, and \( \cos(\cdot, \cdot) \) denotes the cosine similarity between two vectors. To balance precision and recall, we compute the $F_1$ score as harmonic mean of them.

\subsection{Evaluation Dataset}

One prerequisite assumption of EmoCue-360 is that the human-annotated data used for evaluation must be of sufficiently high quality and capable of covering all major emotional cues present in the video segments. At present, among publicly available datasets, only the EMER dataset~\cite{lian2023explainable} can reasonably meet this requirement in terms of annotation quality and cue-level granularity. To further facilitate systematic evaluation research in the EMER task, we additionally construct a new benchmark, EmoCue-Eval, which contains 400 carefully and finely annotated test samples covering diverse scenarios, speakers, and emotional states. To the best of our knowledge, EmoCue-Eval is currently the largest EMER benchmark of its kind. This dataset provides a more systematic benchmark for future research on the comparison of EMER models.

\begin{table*}
  \centering
  \renewcommand{\arraystretch}{1.1}
  \caption{Performance comparison on EMER and EmoCue-Eval benchmarks (A: audio, V: video). Metrics include BLEU, CIDEr, and EmoCue-360 scores for visual (Vis-Emo), audio (Aud-Emo) and global (Glo-Emo) emotional cues with precision (P) and recall (R), where Mean is the average of all P/R scores. \textbf{Bold} indicates best results and \underline{underline} indicates second-best. }
  \begin{tabular}{l|c|c|ccc|*{5}{c|}c|c}
    \hline
    \multirow{2}{*}{Model} & \multirow{2}{*}{Modal} & \multirow{2}{*}{Size \textdownarrow}  & \multirow{2}{*}{$\text{BLEU}_1$ \textuparrow} & \multirow{2}{*}{$\text{BLEU}_4$ \textuparrow} & \multirow{2}{*}{$\text{CIDEr}$ \textuparrow} & \multicolumn{2}{c|}{Vis-Emo} & \multicolumn{2}{c|}{Aud-Emo} & \multicolumn{2}{c}{Glo-Emo} & \multirow{2}{*}{Mean \textuparrow}\\
    & & & & & & P \textuparrow & R \textuparrow & P \textuparrow & R \textuparrow & P \textuparrow & R \textuparrow\\
    \hline
    \multicolumn{13}{c}{\textbf{EMER}}\\
    \hline
    \textbf{General MLLMs} & & & & & & & & & & & \\
    Qwen2-Audio~\cite{chu2024qwen2} & A & 7B & 0.0848 & 0.0086 & 0.0004 & - & - & 38.4 & 41.8 & 47.9 & 42.5 & 42.6 \\
    SALMONN~\cite{tang2023salmonn} & A & 7B & 0.2297 & 0.0170 & 0.0018 & - & - & \underline{43.5} & \underline{48.4} & 45.2 & 41.2 & 44.6 \\
    LLaVA-NeXT~\cite{li2024llava} & V & 7B & 0.2208 & 0.0144 & \underline{0.0072} & \underline{48.8} & 46.2 & - & - & 46.7 & 44.8 & \underline{46.6} \\
    Qwen2.5-VL~\cite{bai2025qwen2} & V & 3B & 0.1950 & 0.0234 & 0.0075 & 42.0 & 33.7 & - & - & 48.7 & 46.4 & 42.7 \\
    LLaMA-VID~\cite{li2024llama} & V & 7B & 0.2165 & 0.0311 & 0.0034 & 47.7 & \underline{49.1} & - & - & 45.4 & 42.2 & 46.1\\
    mPLUG-Owl~\cite{ye2023mplug} & V & 7B &0.2175 & 0.0207 & 0.0037 & 43.8 & 47.9 & - & - & 45.3 & 43.0 & 45.0\\
    Video-ChatGPT~\cite{maaz2023video} & V & 7B & 0.2478 & 0.0369 & 0.0023 & 45.1 & 46.7 & - & - & 45.0 & 41.9 & 44.7\\
    VideoChat~\cite{li2023videochat} & V & 7B & \underline{0.2542} & 0.0254 & 0.0055 & 42.6 & 48.6 & - & - & 42.4 & 42.7 & 44.1\\
    VideoChat2~\cite{li2024mvbench} & V & 7B & 0.1251 & 0.0155 & 0.0001 & 41.4 & 36.8 & - & - & 46.7 & 39.2 & 41.0\\
    \hline
    \textbf{Emotional MLLMs} & & & & & & & & & & & \\
    SECap~\cite{xu2024secap} & A & 7B & 0.0001 & 0.0000 & 0.0000 & - & - & 28.7 & 29.4 & 43.0 & 35.4 & 34.1\\
    Emotion-LLaMA~\cite{cheng2024emotion} & V+A & 7B & 0.1758 & 0.0289 & 0.0017 & \textbf{51.1} & 39.1 & 39.2 & 46.7 & 49.5 & 46.9 & 45.4\\
    AffectGPT~\cite{lian2025affectgpt} & V+A & 7B & 0.2418 & \textbf{0.0691} & 0.0029 & 34.2 & 27.5 & \textbf{46.1} & 46.8 & \textbf{50.9} & \underline{48.0} & 42.3\\
    XEmoGPT (ours) & V+A & 4B & \textbf{0.2965} & \underline{0.0655} & \textbf{0.0349} & 44.7 & \textbf{51.7} & 38.2 & \textbf{50.3} & \underline{50.4} & \textbf{49.6} & \textbf{47.5}\\
    \hline
    \multicolumn{12}{c}{\textbf{EmoCue-Eval}}\\
    \hline
    \textbf{Emotional MLLMs} & & & & & & & & & & \\
    SECap~\cite{xu2024secap} & A & 7B & 0.0001 & 0.0000 & 0.0000 & - & - & 26.5 & 22.1 & 39.1 & 30.6 & 29.6 \\
    Emotion-LLaMA~\cite{cheng2024emotion} & V+A & 7B & 0.1326 & \underline{0.0102} & 0.0000 & \textbf{49.9} & \underline{39.5} & 37.4 & \underline{36.7} & \underline{41.7} & \underline{37.8} & \underline{40.5}\\
    AffectGPT~\cite{lian2025affectgpt} & V+A & 7B & \underline{0.1368} & 0.0076 & 0.0000 & 32.7 & 25.1 & \underline{42.1} & 34.9 & 39.2 & 35.9 & 35.1\\
    XEmoGPT (ours) & V+A & 4B & \textbf{0.2430} & \textbf{0.0488} & \textbf{0.0003} & \underline{43.2} & \textbf{48.7} & \textbf{49.2} & \textbf{49.3} & \textbf{45.1} & \textbf{41.6} & \textbf{46.2}\\
    \hline
  \end{tabular}
\label{tab:exp1}
\end{table*}

\section{Experiments}

We conduct a series of experiments to evaluate the effectiveness of XEmoGPT and to address the following research questions:
\begin{itemize}
    \item \textbf{RQ1}: How does XEmoGPT perform on the EMER task compared with existing methods?
    \item \textbf{RQ2}: How robust is EmoCue-360 to variations in prompt templates and text styles?
    \item \textbf{RQ3}: Can VECB and AECB enhance the emotional cue perception of general-purpose modality encoders?
    \item \textbf{RQ4}: Can VECB and AECB improve XEmoGPT’s ability for emotional cue reasoning?
    \item \textbf{RQ5}: Is the quality of the EmoCue-Instruct dataset sufficient to support model training?
\end{itemize}

\subsection{Implementation Details}

Our experiments were conducted on a machine equipped with two NVIDIA A100 GPUs (80GB memory each), utilizing mixed-precision training with the BF16 format to improve computational efficiency. For multimodal input processing, each video was uniformly subsampled into 16 representative key frames at a spatial resolution of $224\times224$, followed by CLIP-style normalization. Audio waveforms were processed at a sampling rate of 16 kHz, from which HuBERT features were extracted. The hyperparameter settings for the three training stages are summarized as follows: (1) the VECB module was trained for 30 epochs using a batch size of 512 and a learning rate of $3\times10^{-4}$, with CLIP-ViT~\cite{radford2021learning} visual features kept frozen throughout training; (2) the AECB module was trained under comparable settings, with a batch size of 448 and frozen HuBERT-Chinese~\cite{hsu2021hubert} audio features; and (3) during the instruction fine-tuning stage, we adopted Qwen3-4B~\cite{yang2025qwen3} as the backbone LLM and applied LoRA~\cite{hu2022lora} adaptation with a rank of 512. In this stage, the batch size was reduced to 8 and the learning rate was set to $1\times10^{-5}$.

\subsection{Zero-Shot Evaluation (RQ1)}

To demonstrate XEmoGPT’s strong capabilities in perceiving and reasoning over emotional cues, we conduct zero-shot evaluations on two benchmarks: the EMER dataset and EmoCue-Eval. Evaluation metrics include both conventional text generation metrics such as BLEU~\cite{papineni2002bleu} and CIDEr~\cite{vedantam2015cider}, as well as the proposed EmoCue-360, which assesses emotional cue-level performance.

\begin{figure}[t]
\centering
\includegraphics[width=\columnwidth]{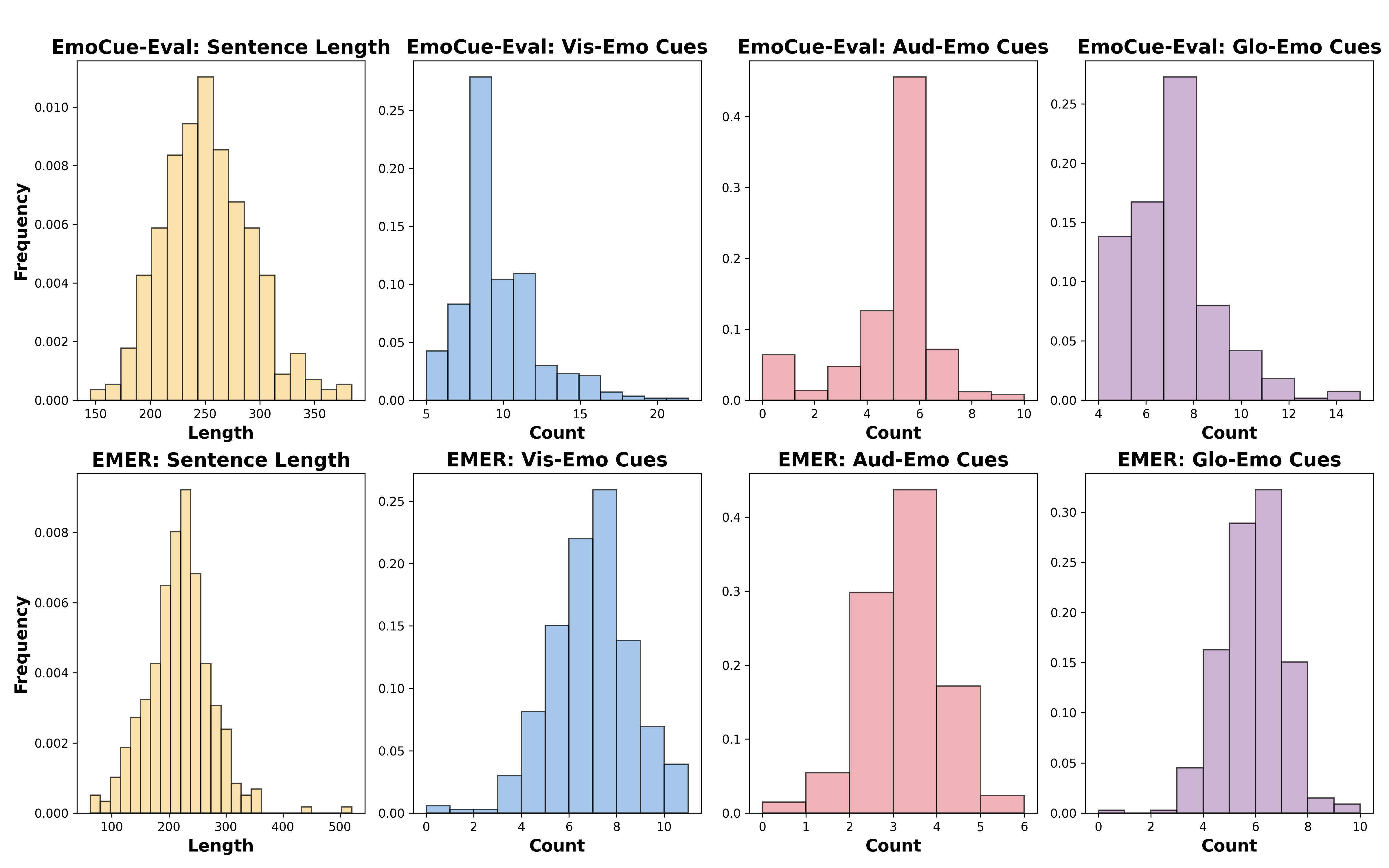}
\caption{Comparison of quality between EmoCue-Eval and EMER datasets, showing annotation length (column 1) and the distribution of emotional cue counts (columns 2–4).}
\label{fig:6}
\end{figure}

As shown in Table~\ref{tab:exp1}, XEmoGPT achieves the highest Mean score on the EMER dataset, outperforming all baseline models. It leads in BLEU-1 and CIDEr, and obtains the highest recall under EmoCue-360 across all emotional cue types. Although its precision scores on EMER are relatively lower, this behavior can be explained by the characteristics of the EMER annotations rather than model deficiencies.

Specifically, Figure~\ref{fig:6} provides a detailed comparison between EMER and EmoCue-Eval in terms of annotation length and emotional cue density. As illustrated in the figure, EMER annotations are significantly shorter and contain fewer explicit emotional cues per sample. In many cases, emotional descriptions in EMER focus primarily on high-level emotional states while omitting fine-grained visual or auditory evidence. As a result, when evaluated using a cue-level metric such as EmoCue-360, XEmoGPT tends to generate additional valid emotional cues that are perceptually grounded but absent from the EMER annotations. These extra cues are consequently counted as false positives, leading to lower precision scores despite being semantically reasonable.

In contrast, EmoCue-Eval provides comprehensive and dense cue annotations across visual, auditory, and global dimensions, substantially reducing annotation sparsity. Under this setting, XEmoGPT consistently outperforms all baseline methods across nearly all metrics, including both precision and recall. Although Emotion-LLaMA slightly surpasses XEmoGPT in visual cue precision, it underperforms in recall and global cue reasoning, indicating a more conservative but less comprehensive explanation strategy.

\subsection{Robustness Analysis of EmoCue-360 (RQ2)}

This experiment evaluates the robustness of the EmoCue-360 metric. Specifically, we assess the robustness from two perspectives: independence from prompt templates and insensitivity to text styles.

\subsubsection{Independence from Prompt Templates}
\begin{figure}[t]
\centering
\includegraphics[width=\columnwidth]{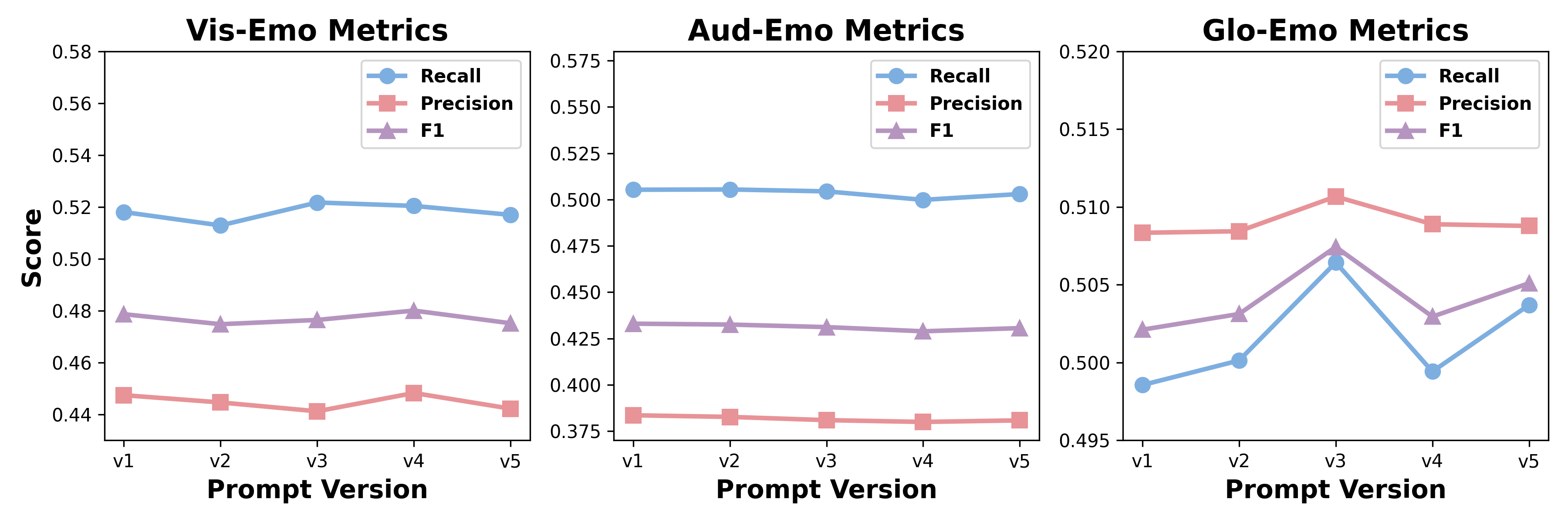}
\caption{EmoCue-360 metric stability across 5 prompt templates, showing consistent performance for visual, audio and global emotional cues.}
\label{fig:5}
\end{figure}

We designed five distinct prompt templates and employed \texttt{deepseek-chat} to extract atomic-level emotional cues from the EMER dataset along three dimensions. Based on the cues generated from each template, we computed EmoCue-360 scores for XEmoGPT. As shown in Figure~\ref{fig:5}, EmoCue-360 demonstrates consistent robustness against variations in prompt design. In particular, Table~\ref{tab:exp3} shows that the range and standard deviation across all evaluation dimensions remain stably at the order of $10^{-3}$.

\begin{table}
  \centering
  \renewcommand{\arraystretch}{1.2}
  \caption{Standard deviation (Std.) and range (Rg.) of precision (P), recall (R), and F1 scores for Vis-Emo (visual), Aud-Emo (audio), and Glo-Emo (global) emotional cues. All values are scaled by $10^{-3}$.}
  \begin{tabular}{c|c|c|c|c|c|c|c|c|c}
    \hline
    & \multicolumn{3}{c|}{Vis-Emo} & \multicolumn{3}{c|}{Aud-Emo} & \multicolumn{3}{c}{Glo-Emo} \\
    & P & R & F1 & P & R & F1 & P & R & F1 \\
    \hline
    Std. \textdownarrow & 3.1 & 3.4 & 2.3 & 1.5 & 2.3 & 1.6 & 0.9 & 2.3 & 2.1 \\
    Rg. \textdownarrow & 7.1 & 8.9 & 5.3 & 3.6 & 5.6 & 4.1 & 2.3 & 5.8 & 5.3 \\
    \hline
  \end{tabular}
  \label{tab:exp3}
\end{table}

\begin{figure}[t]
\centering
\includegraphics[width=\columnwidth]{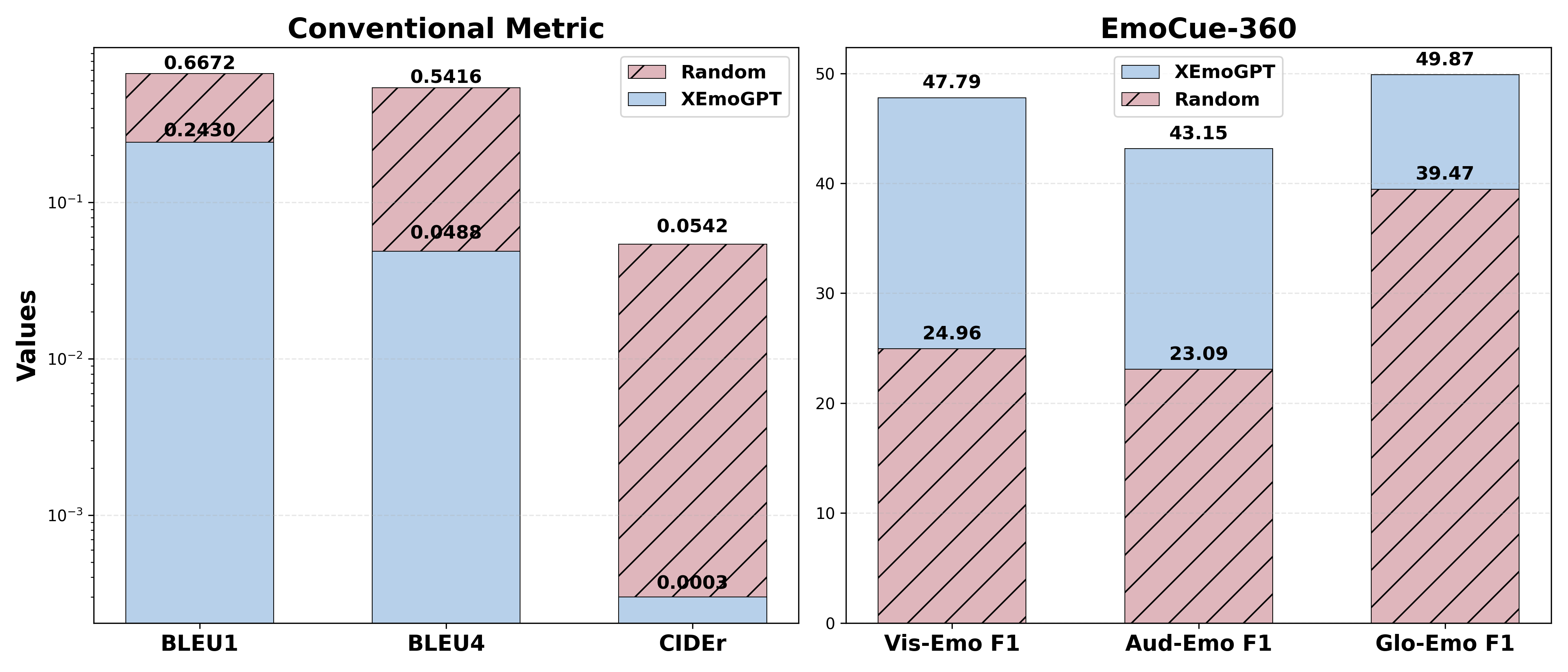}
\caption{Comparison between conventional metrics and EmoCue-360: “Random” represents manually constructed sample pairs, while “XEmoGPT” indicates its scores on the EmoCue-Eval dataset.}
\label{fig:7}
\end{figure}

\subsubsection{Insensitivity to Text Style}
We designed a template and constructed a set of emotional cues covering both visual and audio modalities. A subset of cues was randomly inserted into the template to synthesize two text segments per trial, simulating human annotations and model outputs, respectively, yielding 50 pairs in total. Conventional text generation metrics and EmoCue-360 were then computed for these pairs. As shown in Figure~\ref{fig:7}, conventional metrics yield consistently higher scores for the synthesized samples than XEmoGPT’s performance on EmoCue-Eval. In contrast, under EmoCue-360, XEmoGPT achieves consistently higher scores than the synthesized samples. This indicates that, compared with conventional metrics, EmoCue-360 places greater emphasis on the consistency of emotional cues rather than the overall stylistic similarity of the generated text.

\subsection{Comparison with General Encoders (RQ3)}

To verify the effectiveness of the proposed VECB and AECB modules in enhancing the emotional cue perception capabilities of general-purpose modality encoders, we conduct evaluations on two representative emotion recognition benchmarks: M3ED and MER2023 Track1. These benchmarks differ substantially in data distribution, annotation protocols, and modality characteristics, thereby providing a comprehensive testbed for assessing the generalization ability of cue-enhanced encoders. We report both Unweighted Average Recall (UAR) and Weighted Average Recall (WAR) to account for class imbalance and overall recognition accuracy.

As shown in Table~\ref{tab:exp2}, integrating VECB and AECB consistently leads to significant performance improvements over the corresponding backbone encoders. Compared with strong general-purpose baselines such as CLIP, LongCLIP, and VideoCLIP-XL for the visual modality, CLIP+VECB achieves superior performance across both datasets, indicating that explicitly modeling emotional cues yields more discriminative representations than scaling model size or extending temporal context alone. In particular, CLIP+VECB attains the best overall performance on MER2023, achieving a UAR of 42.6 and a WAR of 43.8.

\begin{table}
  \centering
  \renewcommand{\arraystretch}{1.2}
  \caption{Comparison of modality encoder performance on M3ED and MER2023 Track1 using UAR and WAR. \textbf{Bold} denotes best, \underline{underline} denotes second-best.}
  \begin{tabular}{l|c|c|c|c|c}
    \hline
    \multirow{2}{*}{Model} & \multirow{2}{*}{Modal} & \multicolumn{2}{c|}{M3ED} & \multicolumn{2}{c}{MER2023} \\
     & & UAR & WAR & UAR & WAR \\
    \hline
    CLAP~\cite{wu2023large} & A & 18.3 & 4.5 & 17.5 & 14.1 \\
    LanguageBind~\cite{zhu2024languagebind} & A & 17.5 & 16.6 & 19.6 & 10.9 \\
    CLIP~\cite{radford2021learning} & V & 22.6 & 23.7 & 30.3 & 29.7 \\
    LanguageBind~\cite{zhu2024languagebind} & V & 16.4 & 11.7 & 28.1 & 28.0 \\
    LongCLIP~\cite{zhang2024long} & V & 23.0 & 23.8 & 39.1 & 41.1 \\
    VideoCLIP-XL~\cite{wang2024videoclip} & V & 20.7 & 24.7 & 32.6 & 35.0 \\
    \hline
    HuBERT+AECB & A & \textbf{26.4} & \underline{32.2} & \underline{40.4} & \underline{43.6} \\
    CLIP+VECB & V & \underline{24.8} & \textbf{33.0} & \textbf{42.6} & \textbf{43.8} \\
    \hline
  \end{tabular}
\label{tab:exp2}
\end{table}

For the auditory modality, HuBERT equipped with AECB substantially outperforms audio-only baselines such as CLAP and LanguageBind. Notably, HuBERT+AECB achieves the highest UAR on M3ED, demonstrating that cue-aware audio representations are especially effective in scenarios where emotional signals are subtle and distributed across acoustic patterns. The competitive performance on MER2023 further suggests that AECB generalizes well across datasets with different recording conditions and label distributions.

These results highlight an important distinction between general-purpose multimodal encoders and cue-enhanced encoders. While existing encoders excel at capturing global semantic information, they are not explicitly optimized for emotionally salient patterns. By introducing VECB and AECB, the encoders are guided to attend to fine-grained, emotionally relevant regions and acoustic characteristics, leading to more robust and transferable emotional representations.

\begin{table}
  \centering
  \renewcommand{\arraystretch}{1.2}
  \caption{Ablation results of the modality-specific bridging modules. “Vis-Emo,” “Aud-Emo,” and “Glo-Emo” denote the F1 scores for visual, audio, and global emotional cue, respectively. VECB and AECB refer to the Visual and Audio Emotional Cue Bridge modules, which enhance modality-specific emotional cue reasoning for XEmoGPT.}
  \begin{tabular}{cc|ccc}
    \hline
    \multicolumn{2}{c|}{Module} & Vis-Emo & Aud-Emo & Glo-Emo \\
    \hline
    VECB & AECB & F1 \textuparrow & F1 \textuparrow & F1 \textuparrow \\
    \hline
    - & - & 46.941 & 42.325 & 47.853 \\
    \checkmark & - & \underline{47.582} & 42.465 & \underline{48.764} \\
    - & \checkmark & 46.361 & \underline{43.125} & 48.372 \\
    \checkmark & \checkmark & \textbf{47.785} & \textbf{43.148} & \textbf{49.866} \\
    \hline
  \end{tabular}
  \label{tab:exp4}
\end{table}

\begin{figure*}[t]
\centering
\includegraphics[width=0.95\textwidth]{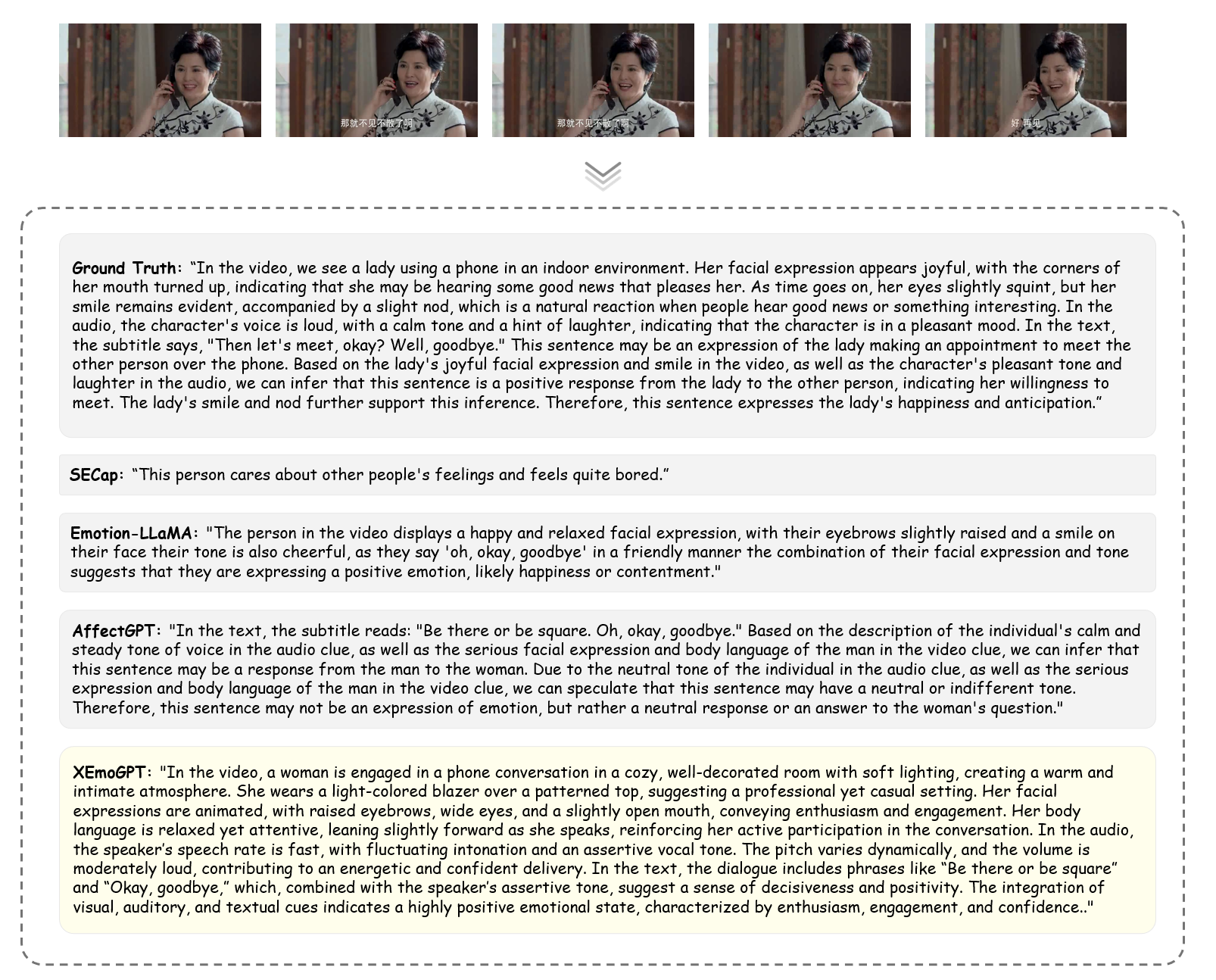}
\caption{Case Study on EMER: Sample Comparison for sample\_00000007}
\label{fig:9}
\end{figure*}

\subsection{Ablation Study (RQ4)}

Table~\ref{tab:exp4} presents an ablation study evaluating the individual and combined contributions of the modality-specific bridging modules. The results reveal several important insights into how VECB and AECB facilitate emotional cue perception and reasoning in XEmoGPT.

First, each bridging module exhibits clear modality-specific effectiveness. Introducing VECB leads to a noticeable improvement in visual emotional cue recognition, increasing the Vis-Emo F1 score from 46.941 to 47.582, while leaving the audio cue performance largely unchanged. Similarly, incorporating AECB primarily enhances auditory emotional cue reasoning, raising the Aud-Emo F1 score from 42.325 to 43.125, with only marginal impact on visual cues. These results confirm that the two modules are not generic performance boosters but instead selectively strengthen emotion-related representations within their respective modalities.

Second, when both VECB and AECB are enabled, XEmoGPT achieves the best performance across all three dimensions, including global emotional cues. Notably, the Glo-Emo F1 score improves from 47.853 in the baseline model to 49.866 when both modules are applied. This improvement exceeds the gains obtained by introducing either module alone, suggesting a complementary effect between visual and auditory emotional cues. By enhancing modality-specific cue perception, the model is better equipped to perform holistic emotional reasoning that integrates evidence across modalities.

Overall, these ablation results validate the design motivation of VECB and AECB. By explicitly modeling emotional cues at the modality level, the proposed bridging modules provide a more structured and explainable pathway for emotional reasoning, ultimately leading to more accurate and comprehensive emotion explanations.

These quantitative findings are further supported by qualitative evidence presented in the case study shown in Figure~\ref{fig:9}, where the joint use of VECB and AECB enables XEmoGPT to identify a more complete, coherent, and perceptually grounded set of emotional cues across multiple modalities.

\begin{figure*}[t]
\centering
\includegraphics[width=0.95\textwidth]{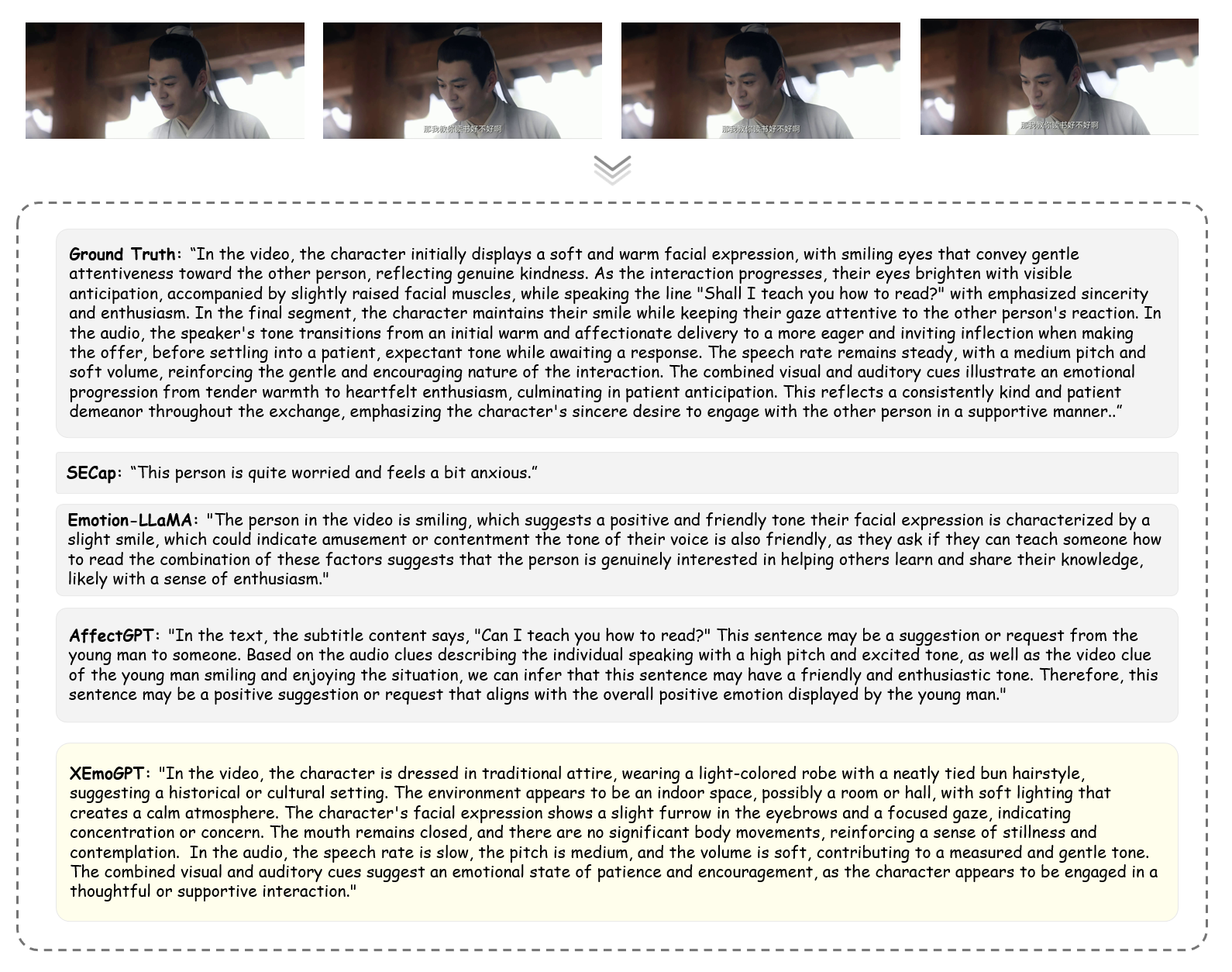}
\caption{Case Study on EmoCue-Eval: Sample Comparison for samplenew3\_00070374}
\label{fig:10}
\end{figure*}

\subsection{Evaluation of EmoCue-Instruct Quality (RQ5)}
\label{sec:eval_emocue_inst}. 

\begin{figure}[t]
\centering
\includegraphics[width=\columnwidth]{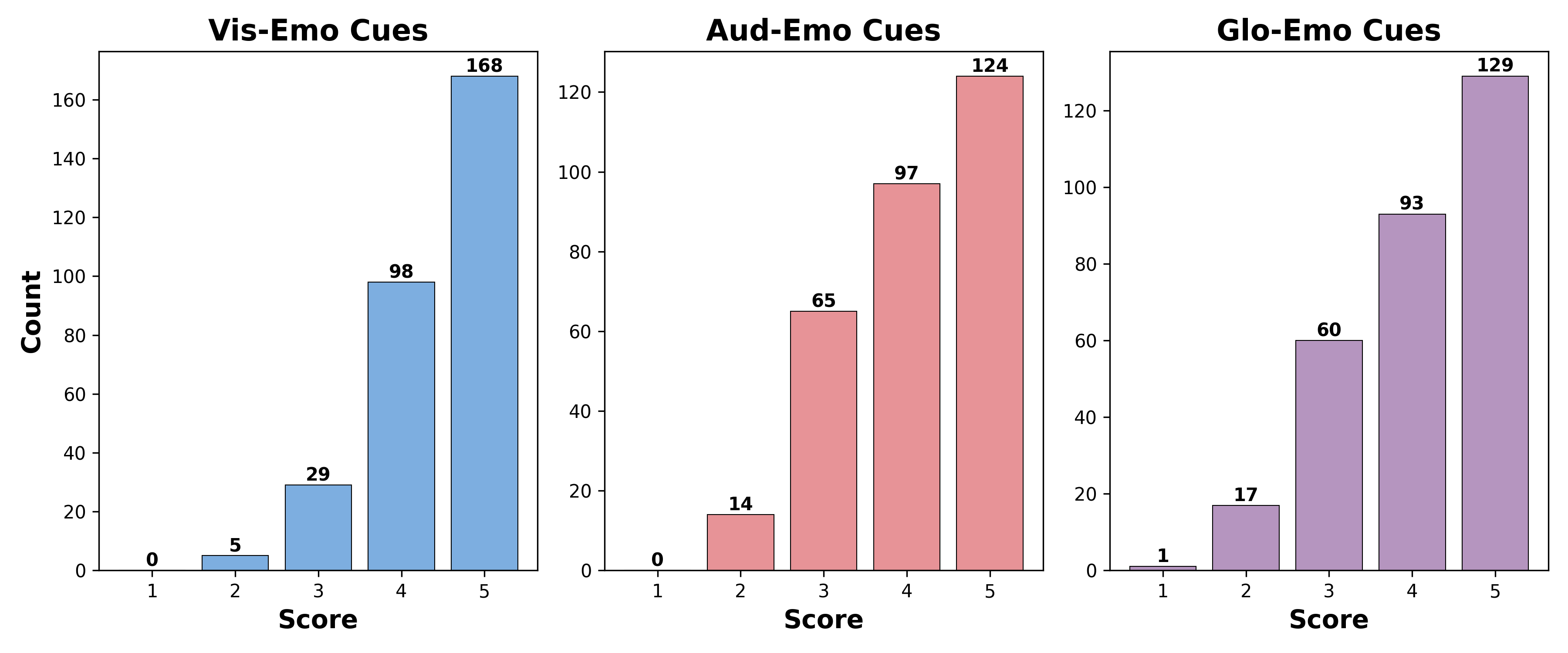}
\caption{Distribution of human evaluation scores for the EmoCue-Instruct dataset. Scores are averaged over three experts across visual, audio, and global emotional cue dimensions. Each score ranges from 1 (worse than MER-Caption+) to 5 (better than MER-Caption+).}
\label{fig:8}
\end{figure}

As discussed in Section~\ref{sec:training_dataset}, the emotional descriptions in the EmoCue-Instruct dataset are primarily generated by MLLMs through reverse inference from emotion labels, followed by summarization using \texttt{DeepSeek-Chat}. Since this process involves no human intervention, a manual evaluation is necessary to assess the annotation quality of the generated data. 

In contrast, the quality of the EmoCue-ShortCaption dataset has been indirectly validated through the strong performance of VECB and AECB in Table~\ref{tab:exp2}. For this reason, and considering the high cost of manual evaluation, we do not conduct additional human assessment for this dataset.

\subsubsection{Evaluation Protocol} 

We randomly sample 300 instances from the EmoCue-Instruct dataset. Since this dataset is adapted from MER-Caption+, we locate the corresponding annotations in MER-Caption+ as human-labeled references. Three experts with research experience in affective computing are invited to score each sample, comparing the quality of annotations between the two datasets. The evaluation focuses on three dimensions: \textit{visual emotional cues}, \textit{audio emotional cues}, and \textit{global emotional cues}. Each dimension is rated on a 1–5 scale, where a score of 1 indicates that the annotation quality of EmoCue-Instruct is inferior to that of MER-Caption+, a score of 3 denotes comparable quality between the two datasets, and a score of 5 indicates that EmoCue-Instruct provides better annotations than MER-Caption+.

As shown in Figure~\ref{fig:8}, EmoCue-Instruct achieves higher average scores across all three dimensions, indicating that its automatically generated annotations maintain high quality and are suitable for training explainable multimodal emotion recognition models.

\subsection{Case Study}

This section presents case studies based on two datasets: EMER and EmoCue-Eval. Figure~\ref{fig:9} illustrates a representative sample from the EMER dataset, while Figure~\ref{fig:10} shows a sample from EmoCue-Eval. Each case includes a video frame and the corresponding subtitles generated by various emotional multimodal language models.

Figure~\ref{fig:9} shows a representative example from EMER. Compared with baseline methods, XEmoGPT provides explanations that are more explicitly grounded in multimodal emotional cues. While SECap produces an incorrect high-level emotional judgment and Emotion-LLaMA and AffectGPT mainly offer coarse-grained emotional summaries, XEmoGPT identifies concrete visual and auditory cues and reasons from these observations to infer a nuanced emotional state, which aligns more closely with the ground truth.

Figure~\ref{fig:10} presents a sample from EmoCue-Eval with dense cue annotations. In this case, XEmoGPT more effectively integrates visual, auditory, and textual cues to infer a positive and engaged emotional state, whereas baseline methods either produce overly generic descriptions or misinterpret the emotional tone. These examples qualitatively demonstrate XEmoGPT’s advantage in cue-level emotional reasoning.

\section{Conclusion}

We introduce XEmoGPT, a novel MLLM that advances explainable emotion reasoning through the perception of emotional cues across modalities. To strengthen cue-level perception, we introduce VECB and AECB, which enhance modality encoders and lay the foundation for multimodal emotional reasoning. We additionally construct the EmoCue dataset to provide fine-grained supervision for cue-level emotional reasoning. To enable explainable evaluation, we develop EmoCue-360 for cue-level assessment, and release EmoCue-Eval, the largest human-annotated benchmark for EMER to date. Experimental results demonstrate that XEmoGPT exhibits strong emotional cues perception and reasoning across modalities. 

Future work will focus on exploring the underlying mechanisms of cue–emotion reasoning, as the current model relies on large amounts of noisy data to learn such reasoning chains. A key direction is to investigate how to effectively filter out spurious cues and achieve more reliable cue–emotion reasoning. Potential directions include improving the sentence-level text encoder of EmoCue-360 by employing an emotion-specific encoder to achieve more sensitive and context-aware cue matching, as well as optimizing the annotation strategies of the EmoCue dataset to further improve the quality and consistency of training supervision.

\bibliographystyle{IEEEtran}
\bibliography{ref}

\end{document}